\documentclass[12pt]{article}
\usepackage{amsfonts}

\def\addcite#1{[???]}
\def\chisb{\raise0.2em\hbox{$\chi$}SB}

\def\gray{\special{ps: 0.4 setgray}}
\def\black{\special{ps: 0.0 setgray}}

\newcommand{\draft}{
\newcount\timecount
\newcount\hours \newcount\minutes  \newcount\temp \newcount\pmhours

\hours = \time
\divide\hours by 60
\temp = \hours
\multiply\temp by 60
\minutes = \time
\advance\minutes by -\temp
\def\hour{\the\hours}
\def\minute{\ifnum\minutes<10 0\the\minutes
            \else\the\minutes\fi}
\def\clock{
\ifnum\hours=0 12:\minute\ AM
\else\ifnum\hours<12 \hour:\minute\ AM
      \else\ifnum\hours=12 12:\minute\ PM
            \else\ifnum\hours>12
                 \pmhours=\hours
                 \advance\pmhours by -12
                 \the\pmhours:\minute\ PM
                 \fi
            \fi
      \fi
\fi
}
\def\fullclock{\hour:\minute}
\begin{centering}
\gray
\special{ps: -90 rotate}
\special{ps: -5100 -5000 translate}
\font\Hugett  =cmtt12 scaled\magstep3
{\Hugett Draft: \today,\clock}
\black
\special{ps: 90 rotate}
\special{ps: 5000 -5100 translate}
\end{centering}
\vskip -1.7cm
$\phantom{a}$
} 
\catcode`\@=11 
\def\lsim{\mathrel{\mathpalette\@versim<}}
\def\gsim{\mathrel{\mathpalette\@versim>}}
\def\@versim#1#2{\vcenter{\offinterlineskip
        \ialign{$\m@th#1\hfil##\hfil$\crcr#2\crcr\sim\crcr } }}
\catcode`\@=12 

\def\nextline{\hfill\break}

\def\mycomm#1{\nextline\strut\kern-3em{\tt ====> #1}\nextline}

\def\nextline{\hfill\break}

\newcommand{\beq}{\begin{equation}}
\newcommand{\eeq}{\end{equation}}
\newcommand{\bea}{\begin{eqnarray}}
\newcommand{\eea}{\end{eqnarray}}
\newcommand{\Seff}{S_{\kern-0.1em\hbox{\small \it eff}}}
\newcommand{\Leff}{{\cal L}_{\kern-0.1em\hbox{\small \it eff}}}
\newcommand{\SMB}{S_{\kern-0.1em\hbox{\small \it MB}}}
\newcommand{\Smb}{S_{\kern-0.1em\hbox{\small \it m-b}}}

\def\eff{\hbox{\small\it eff}\,}
\def\SeffU{S_{\kern -0.1em \eff}[u]}
\def\dminus{\partial_{-}}
\def\Tr{\,\hbox{Tr}\,}
\def\lsquare{\left[}

\def\mprime{{m^{\prime}}}
\def\mtilde{{\hbox{$\tilde m$}}}
\def\alphaprime{{\alpha^{\prime}}}
\def\chiaj{{\chi_{}}_\aj}
\def\jprime{{j^{\prime}}}
\def\spi{\sqrt{4 \pi}}
\def\aj{{\alpha j}}

\def\eqref#1{(\ref{#1})}

\def\downstrut{\vrule height 0ex depth 0.7ex width 0pt}

\def\chiaj{{\chi_{}}_\aj}

\def\chial{{\chi_{}}_\al}

\def\chibl{{\chi_{}}_\bl}

\def\naj{n_\aj}
\def\nal{n_\al}
\def\spi{\sqrt{4 \pi}}
\def\sspi{\sqrt{\pi}}

\def\ac{{\alpha_c}}
\def\aj{{\alpha j}}

\def\al{{\alpha l}}

\def\bl{{\beta l}}

\def\alphaprime{{\alpha^{\prime}}}

\begin{document}

\begin{flushright}
CERN-TH/2003-114\\
WIS/5/03/May-DPP\\
TAUP--2733-03
\end{flushright}
\vskip0.5cm
\begin{center}
{\Large\bf Meson-Baryon Scattering in QCD$_2$ \\
for any Coupling}
\end{center}
\medskip
\begin{center}
{\bf John Ellis}\footnote{\tt john.ellis@cern.ch}\\
{\em Theory Division, CERN, Geneva, Switzerland}\\
\vspace*{0.5cm}
and 
\\
\vspace*{0.5cm}
{\bf Yitzhak Frishman}\footnote{\tt yitzhak.frishman@weizmann.ac.il}\\
{\em Department of Particle Physics\\
Weizmann Institute of Science, 76100 Rehovot, Israel}\\
\vspace*{0.5cm}
and 
\\
\vspace*{0.5cm}
{\bf Marek Karliner}\footnote{\tt marek@proton.tau.ac.il}\\
{\em School of Physics and Astronomy\\
Raymond and Beverly Sackler Faculty of Exact Sciences\\
Tel Aviv University, Tel Aviv, Israel}\\
\end{center}
\begin{abstract}

Extending earlier work on strong-coupling meson-baryon scattering in
QCD$_2$, we derive the effective meson-baryon action for any value of the
coupling constant, in the large-$N_c$ limit. Colour degrees of freedom
play an important role, and we show that meson-baryon scattering can be
formulated as a relativistic potential problem. We distinguish two cases
that are non-trivial for unequal quark masses, and present the resulting
equations for meson-baryon scattering amplitudes.

\end{abstract}
\vfill
\eject

\section{Introduction}

The problem of deriving resonances from QCD is clearly of importance, both
as a fundamental test of QCD and of specific non-perturbative methods.
Experimentally, this field continues to bring surprises, including
enhancements near ${\bar p} p$ thresholds~\cite{BES,Belle,Rosner}, an
exotic $K^+ N$ resonance~\cite{Kyoto,Russia,KN}, and two new mesons
containing $c$ and ${\bar s}$ quarks~\cite{Aubert:2003fg,CLEO}. These
discoveries demonstrate that non-perturbative QCD is not yet fully
`solved'~\cite{Bardeen:2003kt}, and underline the interest in developing
new non-perturbative methods. The dynamics governing the formation of
resonances in QCD is still imperfectly understood, especially the relation
between the spectrum and chiral symmetry breaking (\chisb), possible
`exotic' non-${\bar q} q$ mesons and non-$q q q$ baryons, and combinations
of light and heavy quarks.

QCD in one space and one time dimension, QCD$_2$, is an interesting
laboratory for studying many of these issues. There is no spontaneous
\chisb~ in two dimensions, but a trace of the phenomenon does exist, in 
the
sense that $m_{PS} \propto \sqrt{e_c \, m_q}$, where $m_{PS}$ is the mass
of the lightest pseudoscalar meson, $e_c$ is the gauge coupling and $m_q$
is the quark mass. This formula for $m_{PS}$ is analogous to the case in
four dimensions, with $e_c$ replacing $\Lambda_{QCD}$.

Motivated by this similarity and the tractability of QCD$_2$, we
investigated in~\cite{Frishman:2002ng} the problem of resonances in the
meson-baryon channel in this model. This question is still relevant
both as a crude model for the real world in four dimensions and also as a
testing ground for non-perturbative methods that might have relevance
there and elsewhere.

In~\cite{Frishman:2002ng} we worked in the strong-coupling limit, $e_c/m_q
\gg 1$, namely in the limit where the QCD$_2$ coupling is much larger than
the quark mass.  This would correspond in QCD$_4$ to quark masses being
small compared to the QCD scale, $m_q/\Lambda_{QCD} \ll 1$, the limit
where explicit \chisb~ is small. Moreover, we worked in the approximation
that the baryons are heavy compared to the mesons, which is justified by
the large-$N_c$ limit in QCD. In this double limit, we found no resonances
in meson-baryon scattering, only elastic scattering. This result suggest 
that quark masses are essential complications in QCD, at least in two 
dimensions.

In the present work we relax one of our two previous assumptions, that of
strong coupling, while retaining the assumption that baryons are heavy.
This discussion of general coupling may equivalently be regarded as
introducing finite quark masses. We show that this generalization
necessitates the re-introduction of colour degrees of freedom, which were
decoupled in the previous strong-coupling limit. The scattering problem
may be formulated as a relativistic potential problem that differs
essentially from the previous strong-coupling, low-mass case.  We derive
the corresponding new equations for meson-baryon scattering amplitudes,
valid for any value of the QCD$_2$ coupling.

In Section 2 we discuss the general formulation for QCD$_2$, in bosonized
form, which is necessary for obtaining the meson-baryon scattering
amplitude for an arbitrary value of the coupling. We use this formulation
in Section 3 to derive the effective meson-baryon action, for arbitrary
coupling in the heavy-baryon limit.  We show that in Section 4 that, for
light mesons and heavy baryons, the scattering problem can be transformed
to problem with relativistic potentials, and we evaluate that potential
in two cases that are non-trivial for unequal quark masses.  Section 5
contains our derivation of the equations for scattering amplitudes, and
Section 6 discusses the prospects for future work.

\section{General Formulation}

We start from the discussion of~\cite{Ellis:1992wu}, which presented an
effective action for bosonized QCD$_2$ with $N_f$ flavours and $N_c$ 
colours:
\beq
\SeffU = S_0[u] + 
{e_c^2 N_f \over 8 \pi^2}
\int d\,^2 x \Tr \lsquare \dminus^{-1} \left( u \,\dminus
u^\dagger\right)_c \right]^2
+ \mprime^2 N_\mtilde \int d\,^2 x \Tr \left( u + u^\dagger\right),
\label{eqI}
\eeq
where the traces are over both colour and flavour. The second term is the
result of integrating out the gauge fields, which was made possible by the
quadratic dependence on the gauge fields in two dimensions in the 
light-cone gauge, which we chose. Finally, $S_0[u]$ is the Wess-Zumino-Witten
(WZW) action, representing free fermions.

The above action is exact for any value of the gauge coupling $e_c$. The
matrices $u$ represent the bosonic version of the quark bilinears,
symbolically $\Psi \otimes \bar\Psi$, with both colour and flavour
indices. The symmetry groups are $SU(N_c)$ and $U(N_f)$, for colour and
flavour, respectively, including the $U(1)$ of baryon number, and
\beq
\mprime^2 = m_q C \mtilde,
\label{mprime}
\eeq
where $m_q$ is the quark mass, $\mtilde$ the normal-ordering scale, and $C
\equiv {1 \over 2} e^{\gamma} \approx 0.891$. For a detailed discussion,
see~\cite{Frishman:1992mr}. For the time being, we take all quarks to have
the same mass.

When using the product scheme for bosonization
\beq
{u = h \times g \equiv 
[\,\,\underbrace{SU(N_c)\downstrut}\,\,]_{\lower0.5em\hbox{$N_F$}}\atop
\kern4.1em h}
{\times\atop\phantom{A}}
{[\,\,\underbrace{U(N_F)\downstrut}\,\,]_{\lower0.5em\hbox{$N_c$}}\atop
\kern-1.4em g}
\eeq
and taking the strong-coupling limit ${e_c/ m_q} \to \infty$, one can
eliminate the second term in (\ref{eqI}), after first choosing an
appropriate normal-ordering scale $\tilde m$, as described
in~\cite{Frishman:1992mr}. The colour degrees of freedom $h$ are
completely eliminated in this limit, and one gets an effective action in
terms of flavour degrees of freedom only:
\bea
\tilde\Seff\,[g] &=& N_c S_0[g] + m^2 N_m \int d^2 x 
\left( \, \hbox{Tr}_f \, g + \, \hbox{Tr}_f \, g^\dagger \right),
\label{effective}
\\
\phantom{a}\nonumber\\
{\rm where}\qquad\ m &=& \left[ N_c C m_q \left({e_c \sqrt{N_f}\over
\sqrt{2\pi}}\right)^{\Delta_c} \right]^{1\over 1 + \Delta_c},
\\
\phantom{a}\nonumber\\
\Delta_c &=& { N_c^2 -1 \over N_c (N_c + N_F)}\,\,.
\eea
Note that here the traces are over flavour only.

As shown in~\cite{Frishman:2002ng}, this action results in elastic
scattering only, and is therefore unsuitable for modelling QCD$_4$
realistically. Scattering is a soluble problem in this approximation, with
the transmission and reflection coefficients $T$ and $R$ given
in~\cite{Frishman:2002ng} for all meson energies.

When dealing with finite coupling, we should not use the product scheme,
but rather the `$u$-scheme', where the original matrices $u \in
U(N_F\times N_c)$. The two schemes are equivalent only when the quark
masses are zero, or equivalently in the strong-coupling limit. One
important aspect of the difference, at finite mass, is that the
`$u$-scheme' admits multi-soliton solutions~\cite{Ellis:1992wu} that do
not exist in the product scheme. Each of the solitons in these solutions
carries both colour and flavour, yet the total multi-soliton solutions are
colour-neutral.  These individual coloured solitons were interpreted
in~\cite{Ellis:1992wu} as constituent quarks. When the quark masses are
equal, the solutions can be obtained analytically, and are given by the
sine-Gordon profile. For unequal quark masses, the solutions were obtained
numerically.

\section{Extension to Arbitrary Coupling}

We first consider the second term of (\ref{eqI}), for arbitrary coupling:
\beq
{e_c^2 N_f \over 8 \pi^2}
\int d\,^2 x \Tr \lsquare \dminus^{-1} \left( u \,\dminus
u^\dagger\right)_c \right]^2
\label{eqIII}
\eeq
where $\left( u \,\dminus u^\dagger\right)_c$ is the colour part of 
$M \equiv u \partial_- u^\dagger$, to be computed as
\beq
M_c = \hbox{Tr}_f M - 1/N_c \hbox{Tr}_{f\&c} M,
\label{traceMc}
\eeq
with details to be found in~\cite{Frishman:1992mr}. As already mentioned,
this term represents the interactions, as it arises from integrating out
the gauge potentials. However, we will see that, for the physical
situation we discuss, this term does not contribute to meson-baryon
scattering for any coupling. As a result, the latter is described by the
effective action $\tilde\Seff\,[u]$, whereas in the strong coupling limit
it is described by $\tilde\Seff\,[g]$.

In order to describe meson-baryon scattering, we follow the 
four-dimensional example~\cite{Mattis:1984dh},
and take $u$ to be of the form
\beq
u= \exp(-i \Phi_c) \exp (-i \delta \Phi),
\label{def-u}
\eeq
corresponding to a classical soliton $\Phi_c$ representing one of the
baryons described in~\cite{Ellis:1992wu}, and a small fluctuation $\delta
\Phi$ around it, representing the meson. The resulting action is then
expanded to second order in $\delta \Phi$, yielding a linear equation of
motion for $\delta \Phi$ in the soliton background. The latter serves as
an external potential in which the meson is propagating.

We start by evaluating
\bea
M &\equiv &u \partial_- u^\dagger =\\
&=& \exp (-i \Phi_c)\, \partial_- (\exp i \Phi_c)
 + \exp( -i \Phi_c) \exp (-i \delta \Phi) 
\, [\partial_- \exp (i \delta \Phi)] \exp (i \Phi_c) ,
\nonumber
\label{defM}
\eea
and obtain the equations of motion for the meson field by varying with 
respect to $\delta \Phi$. The variation of (\ref{eqIII})
with respect to $\delta \Phi$ is proportional to 
\beq
{\delta M_c \over \delta(\delta \Phi)}\, 
\partial^{-2} M_c.
\eeq
To compute its variation with respect to $\delta \Phi$, we need only 
the second term $M_2$ of $M$, as the first term $M_1$ 
is independent of $\delta \Phi$.

We take for the soliton a diagonal ansatz, following~\cite{Ellis:1992wu}:
\bea
[\exp (-i \Phi_c)]_{\alpha\alphaprime j\jprime} & =&
\delta_{\alpha\alphaprime}\,\delta_{j\jprime}
\exp{(\displaystyle - i\spi\chiaj)}: \nonumber\\ 
\alpha &=&1,\dots,N_c,\\
j&=&1,\dots,N_f,
\nonumber
\eea
so that
\bea
\left\{ \exp( -i \Phi_c) [ \exp (-i \delta \Phi)  
\,\partial_-\exp (i \delta \Phi) ] 
\exp( i \Phi_c) \right\}_{\alpha j, \alpha^\prime j^\prime} 
   = 
\nonumber
\\
\exp( -i \sqrt{4 \pi} \chi_{\alpha j}) 
\,[ \exp( -i \delta \Phi) 
\,\partial_- \exp (i \delta \Phi)]_{\alpha j, \alpha^\prime j^\prime} 
\,\exp (i \sqrt {4 \pi} \chi_{\alpha^\prime j^\prime}).
\eea
The part of $M$ that contributes to the effective action 
is its colour projection (\ref{traceMc}).
We note that $\hbox{Tr}_{f\&c} M_2 = 0$, and thus
\beq
{[ (M_2)_c ]}_{\alpha, \alpha^\prime} = \sum_j
\exp(-i \sqrt{4 \pi} \chi_{\alpha j}) 
[\exp(-i \delta \Phi) 
\,\partial_-\exp (i \delta \Phi)] _{\alpha j, {\alpha}^\prime j} 
\exp( i \sqrt{4 \pi} \chi_{{\alpha}^\prime j}).
\eeq
The mesons $\delta \Phi$ have to be diagonal in colour, so
\beq
{[(M_2)_c]}_{\alpha, \alpha^\prime}= 
\sum_j[\,\exp (-i \delta \Phi)\, 
\partial_-\exp (i \delta \Phi)\,]_ {\alpha j, \alpha j}
   \delta_{\alpha, \alpha^\prime}.
\eeq
We recall that the flavour structure of the mesons is independent of their
colour indices, and restrict our attention to mesons that have no $U(1)$
flavour part. In this way, we may be sure that classical solutions lead to
stable particles, since their non-vanishing flavour quantum numbers put
them in a different sector from the vacuum. We then have
\beq
\sum_j 
[\, \exp (-i \delta \Phi)  
\,\partial_- \exp (i \delta \Phi)\,]_{\alpha j, \alpha j} = 0,
\eeq
as advertized earlier, and the effective meson-baryon action is 
\bea
\tilde\Smb [\delta \Phi] &=& S_0[u] + m^2 N_m \int d^2 x 
\left( \, \hbox{Tr}\, u + \, \hbox{Tr}\, u^\dagger \right),
\label{MB}
\eea
with $u$ depending on $\delta \Phi$ for fixed $\Phi_c$ as in 
(\ref{def-u}).

\section{Evaluation of the Potential}

The equation of motion for $\delta \Phi$ is obtained
from (\ref{MB}), by first varying with respect to $u$ 
and then varying $u$ with respect to $\delta \Phi$.
To first order in $\delta \Phi$, we find
\beq
\delta u = -i [\exp (-i \Phi_c)] \delta \Phi .
\eeq
The resulting equation of motion is then
\beq
{1\over 4 \pi} \partial_+
\left[ \left(\partial_-u\right) u^\dagger\right]
+  \left( u m^2 - m^2 u^\dagger\right) = 0 ,
\eeq
where $m$ is the diagonal mass matrix: $m=\delta_{ij} m_j$ with (possibly
different) entries $m_j$ corresponding to flavours $j$.
We note that there is the possibility of an 
overall scale ambiguity in $m$, since, when the masses are different, 
there is a question which normal-ordering scale to use, as discussed in 
the Appendix of~\cite{Ellis:1992wu}.
The resulting equation of motion for $\delta \Phi$ is 
\bea
\Box \delta\Phi - i \left( \partial_+\Phi_c \right)
\left( \partial_-\delta\Phi \right)
+i\left(\partial_-\delta\Phi\right)\left(\partial_+\Phi_c\right)
+
\nonumber
\\
\displaystyle{1\over2}
\left[
\delta\Phi \mu^2 \exp(-i\Phi_c)
+\exp(i\Phi_c) \mu^2 \delta\Phi
\right]
=0,
\label{eq-mot}
\eea
where $\mu \equiv m \sqrt{8 \pi}$.

As discussed before, both $\Phi_c$ and $\delta\Phi$ are diagonal in
colour.  Moreover, $\Phi_c$ is diagonal in flavour too. So, taking the
$\,\alpha \alpha j {j}^\prime\,$ matrix element of the equation of motion
(\ref{eq-mot}), we find
\bea
\Box {\delta\Phi}_{\alpha j {j}^\prime}
- i{\left( \partial_+\Phi_c \right)}_{\alpha j}
{\left( \partial_-\delta\Phi \right)}_{\alpha j {j}^\prime}
{+i\left(\partial_-\delta\Phi\right)}_{\alpha j {j}^\prime}
{\left(\partial_+\Phi_c\right)}_{\alpha   {j}^\prime} +
\nonumber 
\\
\displaystyle{1\over2}
\{ {\delta\Phi}_{\alpha j {j}^\prime}
{\mu^2}_{{j}^\prime} [\exp(-i\Phi_c)]_{\alpha {j}^\prime} + [\exp(i\Phi_c)]_{\alpha j} {\mu^2}_j 
{\delta\Phi}_{\alpha j {j}^\prime} \} = 0.
\eea
Examining the classical solutions for the quark solitons inside the
baryons as in~\cite{Ellis:1992wu}, we see that, for a given colour index
$\alpha$, there is only one flavour for which $\Phi_c$ is non-zero. We can
now distinguish three cases.

$\bullet$
The first is when an index $\alpha$ and indices $j$ and
$j^{\prime}$ are chosen in such a way that both $(\Phi_c)_{\alpha j}$ and 
$(\Phi_c)_{\alpha {j}^\prime}$ are zero. In such a case,
\bea
\Box {\delta\Phi}_{\alpha j {j}^\prime} + \displaystyle{1\over2} 
[{\mu^2}_j + {\mu^2}_{{j}^{\prime}}]
{\delta\Phi}_{\alpha j {j}^\prime} = 0,
\\
{\rm where} {(\Phi_c)}_{\alpha j} = 0  \ \hbox{and} \ {(\Phi_c)}_{\alpha 
{j}^\prime} = 
0.
\nonumber
\eea
Thus ${\delta\Phi}_{\alpha j {j}^\prime}$ is a free field with squared 
mass given by the average of $m^2_j$ and $m^2_{j^\prime}$ in this case, 
which we do not discuss further.

$\bullet$
The second case is that of $j = {j}^\prime$, with $\alpha$ such that
${(\Phi_c)}_{\alpha j}$ is a quark soliton inside the baryon. In this 
case,
\beq
\Box {\delta\Phi}_{\alpha j j} + 
{\mu^2}_j \cos [{(\Phi_c)}_{\alpha j}]
{\delta\Phi}_{\alpha j j} = 0.
\label{same j}
\eeq
This case is analogous to the case in Section 2 of~\cite{Frishman:2002ng}, 
with the difference that here the potential is 
determined by $\Phi_c$ which, for the case of unequal masses, 
is not necessarily of the sine-Gordon type.

$\bullet$
The third case is when $j$ is different from $j^\prime$, now with one of
the $\Phi_c$ being a soliton and the other vanishing. Taking
${(\Phi_c)}_{\alpha j}$ to be the soliton, we obtain
\beq
\Box {\delta\Phi}_{\alpha j {j}^\prime}
- i{\left( \partial_+\Phi_c \right)}_{\alpha j}
{\left( \partial_-\delta\Phi \right)}_{\alpha j {j}^\prime} +
\displaystyle{1\over2}
\{ {\mu^2}_{{j}^\prime}  + {\mu^2}_j [\exp(i\Phi_c)]_{\alpha j} \}
{\delta\Phi}_{\alpha j {j}^\prime}  = 0,
\label{diff j}
\eeq
where
\centerline {$j^{\prime} \neq j 
 \  \hbox{and}  \ {(\Phi_c)}_{\alpha {j}^\prime} = 0$}.
This case is analogous to Section 4 of~\cite{Frishman:2002ng}, 
but with the same difference as in (\ref{same j}), i.e., that for
unequal masses the soliton is not of the sine-Gordon type.

\section{Evaluation of Meson-Baryon Scattering}

The equations which determine the static solution $(\Phi_c)_{\alpha j}$ 
were derived in~\cite{Ellis:1992wu}. 
For completeness, we include them here too. First one defines
\beq
(\Phi_c)_{\alpha j} = \sqrt{4 \pi} (\chi_c)_{\alpha j},
\eeq
where the $(\chi_c)_{\alpha j}$ are canonical fields, whose
equations of motion are 
$$ \chi^{\prime\prime}_{\aj}
- 4 \ac \left(  \sum_l \chial - {1\over N_c}
\sum_{\bl} \chibl \right)  - 2 \sqrt{4\pi} m_j^2 \sin \spi \chiaj = 0.
$$
Note the extra factor 2 in front of the mass term, as compared with Eq. (22) 
of~\cite{Ellis:1992wu}, due to an error in this reference.

Choosing the boundary conditions $\chi_{\alpha j}(-\infty) = 0$,
we get as constraints for $\chi_{\alpha j}(+\infty)$, denoted hereafter 
simply by $\chi_{\alpha j}$,
\beq
{1\over\sspi} \chiaj  = \naj \qquad\hbox{integers} ,
\eeq
and
\beq
 \sum_l \nal   =  n \qquad\hbox{independent of $\alpha$}.
\eeq
The baryon number~\footnote{In our normalization,
a single quark carries one unit of baryon number.}
associated with any given flavour $l$ is given by
$$ B_l = \sum_\alpha \nal.
$$
Combining the last two equations, we find
$$
B = \sum_l B_l = nN_c
$$ 
for the total baryon number.

We now continue in a similar manner to~\cite{Frishman:2002ng}, starting
with the first non-trivial case (\ref{same j}) identified above. As the 
soliton solutions are such that there is a unique correspondence between
the colour index $\alpha$ and the flavour index $j$, we suppress
$\alpha$ in what follows. Putting
\beq
\delta\Phi_{j j} =  e^{{-}i{\omega}_j t} u_j(x)
\eeq
with
\beq
u_j(x)\mathop{\longrightarrow}_{x\to\infty} e^{i k x},
\label{right}
\eeq
we find
\beq
\omega_j^2 = k^2 +\mu_j^2,
\eeq
and the equation for $u_j(x)$ is 
\beq
u_j^{\prime\prime}(x) + {\omega_j}^2 u_j - {\mu^2}_j 
[ \cos {(\Phi_c)}_{j}] u_j = 0.
\label{u same j}
\eeq
We define the potential $V_j$ for this scattering process via 
\beq
u_j^{\prime\prime}(x) + {\omega_j}^2 u_j - V_j u_j = 0,
\eeq
and find
\beq
V_j = {\mu^2}_j [ \cos {(\Phi_c)}_{j}].
\eeq
In our normalization the outgoing wave has coefficient 1, 
which is more convenient for numerical calculations,
and the wave for $x \,\to\, {-}\infty$ is now 
\beq
u_j(x) = {1\over T_j} \, e^{i k x} + {R_j\over T_j}\, e^{-i k x},
\qquad\qquad x \,\to\, {-}\infty 
\label{left}
\eeq
in this case.

In the second non-trivial case (\ref{diff j}), we put
\beq
\delta\Phi_{j j^{\prime}} =  e^{{-}i{\omega}_{j j^{\prime}} t} 
u_{j j^{\prime}}(x),
\eeq
so that
$$
u_{j {j}^\prime}^{\prime\prime}(x)
- i (\Phi_c)_{j}^\prime(x)
u_{j {j}^\prime}^\prime(x)  +
$$
\beq
+ \{\omega_{j j^{\prime}}^2 + \omega_{j j^{\prime}} (\Phi_c)_j ^\prime(x) -
\displaystyle{1\over2}
\{ {\mu^2}_{{j}^\prime}  + {\mu^2}_j [\exp(i\Phi_c)]_j \}\}
u_{j {j}^\prime}  = 0.
\eeq
To eliminate the first derivative term in $u$, we substitute
\beq
u_{j {j}^\prime} = [\exp({{i\over2}\Phi_c})]_j \,v_{j {j}^\prime}.
\eeq
This results in
$$
v_{j {j}^\prime}^{\prime\prime}(x) + \{\omega_{j j^{\prime}}^2 
+ \omega_{j j^{\prime}} (\Phi_c)_j ^\prime(x) - {\mu^2}_j [\cos(\Phi_c)]_j\}
v_{j {j}^\prime} +
$$
$$
+ {1\over 2}({\mu^2}_j - {\mu^2}_{{j}^\prime})v_{j {j}^\prime} +
$$
$$
+\{ {1\over 4}[(\Phi_c)_j ^\prime(x)]^2 - {1\over 2}{\mu^2}_j (1 - 
[\cos(\Phi_c)]_j)\}
v_{j {j}^\prime} +
$$
\beq
+ {i\over 2} \{{(\Phi_c)_j}^{\prime\prime}(x) - {\mu^2}_j [\sin (\Phi_c)_j]\}
v_{j {j}^\prime} = 0.
\label{final diff j}
\eeq
We note that the last three lines vanish when all the quark masses are 
equal, as then the soliton is a sine-Gordon one. Thus, the scattering 
would then be only elastic, as found in~\cite{Frishman:2002ng}.

The potential of the scattering is defined here via
\beq
v_{j {j}^\prime}^{\prime\prime}(x) + \omega_{j j^{\prime}}^2 v_{j {j}^\prime}
-V_{j {j}^\prime} v_{j {j}^\prime} = 0,
\eeq
so that
$$V_{j {j}^\prime} =  - \omega_{j j^{\prime}} (\Phi_c)_j ^\prime(x) 
+
{\mu^2}_j [\cos(\Phi_c)]_j $$
$$- {1\over 2}({\mu^2}_j - {\mu^2}_{{j}^\prime})$$
$$- \{ {1\over 4}[(\Phi_c)_j ^\prime(x)]^2 - {1\over 2}{\mu^2}_j 
(1 - [\cos(\Phi_c)]_j)\}$$
\beq
- {i\over 2} \{{(\Phi_c)_j}^{\prime\prime}(x) - {\mu^2}_j [\sin (\Phi_c)_j]\}
\eeq
Taking again
\beq
v_{j j^{\prime}}(x)\mathop{\longrightarrow}_{x\to\infty} e^{i k x},
\label{right prime}
\eeq
we get
\beq
\omega_{j j^{\prime}} = {1\over 2}({\mu^2}_j + {\mu^2}_{{j}^\prime}),
\eeq
and the wave for $x \,\to\, {-}\infty$ is
\beq
v_{j j^{\prime}}(x) = {1\over T_{j j^{\prime}}} \, e^{i k x} + 
{R_{j j^{\prime}}\over T_{j j^{\prime}}}\, e^{-i k x},
\qquad\qquad x \,\to\, {-}\infty 
\label{left prime}
\eeq
in this case.

\section{Discussion}

We have shown that meson-baryon scattering in QCD$_2$ in the large-$N_c$ 
limit is non-trivial for non-zero quark masses, and is described by two 
distinct effective potentials when the quark masses are unequal. These 
effective potentials are not of the sine-Gordon type found in previous 
cases, and we expect the scattering amplitudes also to be non-trivial. 
Their calculation will require numerical analysis, that we postpone for a 
future occasion.

Clearly QCD$_2$ is not a complete laboratory for studying non-perturbative
physics in QCD$_4$. However, it is already quite a rich system, and full
QCD can only be richer still. We have already pointed out the existence of
constituent quark solitons in QCD$_2$~\cite{Ellis:1992wu}, and this
numerical analysis may cast light on their importance in scattering, where
the additive quark model has long been an intriguing approximation.

It is intriguing that the introduction of unequal quark masses is an
essential complication. We recall that the light-cone wave functions for
mesons containing unequal quark masses are expected not to be symmetric,
resulting in a net colour field that underlies this effect. One task of
numerical analysis will be to see what new physics this produces in
meson-baryon scattering and possibly in resonant states. We note that two
of the recent new puzzles in non-perturbative QCD concern systems with
unequal quark masses~\cite{Kyoto,Russia,Aubert:2003fg,CLEO}. It would be
hubristic to suggest that the continuation of our QCD$_2$ studies will
cast light on these puzzles, but they will provide some extra motivation
for our future work in this direction.

\section*{Acknowledgments}
The research of one of us (M.K.) was supported in part by a grant from the
United States-Israel Binational Science Foundation (BSF), Jerusalem and
by the Einstein Center for Theoretical Physics at the Weizmann Institute.

\end{document}